\documentclass[11pt]{article}

\def\Journal#1#2#3#4{{#1} {\bf #2}, #3 (#4)}

\def\NPB{{\em Nucl. Phys.} B}
\def\PLB{{\em Phys. Lett.}  B}
\def\PRL{\em Phys. Rev. Lett.}
\def\PRD{{\em Phys. Rev.} D}

\def\be{\begin{equation}}
\def\ee{\end{equation}}
\def\bea{\begin{eqnarray}}
\def\eea{\end{eqnarray}}

\usepackage{hyperref,bm,amsmath,amssymb,a4wide,graphicx,subfigure,wasysym,amsfonts,cite}

\begin{document}

\title{\textbf{New magnetic field instability and magnetar bursts}}

\author{Maxim Dvornikov$^{a,b}$\thanks{maxdvo@izmiran.ru}
\\
$^{a}$\small{\ Pushkov Institute of Terrestrial Magnetism, Ionosphere} \\
\small{and Radiowave Propagation (IZMIRAN),} \\
\small{108840 Troitsk, Moscow, Russia;} \\
$^{b}$\small{\ Physics Faculty, National Research Tomsk State University,} \\
\small{36 Lenin Avenue, 634050 Tomsk, Russia}}

\date{}

\maketitle

\begin{abstract}
We consider the system of massive electrons, possessing nonzero anomalous magnetic moments, which electroweakly interact with background neutrons under the influence of an external magnetic field. The Dirac equation for such electrons is exactly solved. Basing on the obtained solution, we find that a nonzero electric current of these electrons can flow along the magnetic field. Accounting for the new current in the Maxwell equations, we demonstrate that a magnetic field in this system appears
to be unstable. Then we consider
a particular case of a degenerate electron gas, which may well exist in a
neutron star, and show that a seed magnetic field can be
amplified by more than one order of magnitude. Finally we discuss the application of our results for the explanation of
the electromagnetic radiation emitted by magnetars.
\end{abstract}

\section{Introduction}

The problem of the magnetic field instability is important, e.g.,
in the context of the existence of strong astrophysical magnetic fields~\cite{Spr08}.
Besides the magnetohydrodynamics mechanisms for the generation of
astrophysical magnetic fields, recently the approaches based on the
elementary particle physics were proposed. These approaches mainly
rely on the chiral magnetic effect (CME)~\cite{MirSho15}, which
consists in the generation of the anomalous current of massless charged
particles along the magnetic field $\mathbf{J}_{5} = \alpha_{\mathrm{em}} \left( \mu_{\mathrm{R}}-\mu_{\mathrm{L}} \right) \mathbf{B}/\pi$,
where $\alpha_{\mathrm{em}}\approx1/137$ is the fine structure constant
and $\mu_{\mathrm{R,L}}$ are the chemical potentials of right and
left chiral fermions. If $\mathbf{J}_{5}$ is accounted for in the
Maxwell equations, the magnetic field appears to be unstable and can
experience a significant enhancement. The model for the generation of strong magnetic fields in the dense matter of a neutron star (NS) driven by CME under the influence of the electroweak interaction between electrons and neutrons was developed in a series of our works~\cite{DvoSem15a,DvoSem15b,DvoSem15c,Dvo16d,Dvo17}. Other applications of CME for
the generation of astrophysical and cosmological magnetic fields are
reviewed by Kharzeev~\cite{Kha15}.

However, the existence of CME in astrophysical media is questionable.
Vilenkin~\cite{Vil80} and Dvornikov~\cite{Dvo16a} found that $\mathbf{J}_{5}$ can be
non-vanishing only if the mass of charged particles, forming the current,
is exactly equal to zero, i.e. the chiral symmetry is restored. For
the case of electrons the restoration of the chiral symmetry is unlikely
at reasonable densities which can be found in astrophysics~\cite{Rub86}.
The chiral symmetry can be unbroken in quark matter owing to the strong
interaction effects~\cite{BulCar16}. The magnetic fields generation
in quark matter, which can exist in some compact stars, was discussed
by Dvornikov~\cite{Dvo16b,Dvo16c}. Nevertheless this kind of situation
looks quite exotic.

Therefore the issue of the existence of an electric current $\mathbf{J}\sim\mathbf{B}$
for massive particles, which can lead to the magnetic field instability,
is quite important for the explanation of astrophysical magnetic fields.
One of the examples of such a current in electroweak matter was proposed
by Semikoz \& Sokoloff~\cite{SemSok04}. However, the model developed by Semikoz \& Sokoloff~\cite{SemSok04}
implies the inhomogeneity of background matter. This fact imposes
the restriction on the scale of the magnetic field generated.

In the present work, we discuss another scenario for the magnetic
field instability. It involves the consideration of the electroweak
interaction of massive fermions with background matter along with
nonzero anomalous magnetic moments of these particles. Note that the
electroweak interaction implies the generic parity violation which
can provide the magnetic field instability. Recently, the interpretation
of CME in terms of an effective magnetic moment was considered by Kharzeev et al.~\cite{KhaSteYee17}.

This work is organized as follows. In Sec.~\ref{sec:DIREQ}, we discuss the Dirac equation
for a massive electron with a nonzero anomalous magnetic moment, electroweakly
interacting with background matter under the influence of an external
magnetic field. Then, we describe the main steps in finding the exact solution of this Dirac equation which was previously obtained by Balantsev et al.~\cite{BalStuTok12}.
Using this solution, in Sec.~\ref{sec:CURR}, we calculate the electric current of these electrons along
the magnetic field direction. This current turns out to be nonzero.
Then we consider a particular situation of a strongly degenerate electron
gas, which can be found inside NS. Finally, in Sec.~\ref{sec:INST}, we apply
our results for the description of the amplification of the magnetic
field in NS and briefly discuss the implication of our findings for the
explanation the electromagnetic radiation of compact stars.

\section{Solution of the Dirac equation\label{sec:DIREQ}}

Let us consider an electron with the mass $m$ and the anomalous magnetic
moment $\mu$. This electron is taken to interact electroweakly with
nonmoving and unpolarized background matter consisting of neutrons
and protons under the influence of the external magnetic field along
the $z$-axis, $\mathbf{B}=B\mathbf{e}_{z}$. Accounting for the forward
scattering off background fermions in the Fermi approximation, the
Dirac equation for the electron has the form,
\be\label{eq:Direq}
  \left\{
    \gamma_{\mu}P^{\mu}-m-\mu B\Sigma_{3}-\gamma^{0}
    \left[
      V_{\mathrm{R}}
      \left(
        1+\gamma^{5}
      \right) +
      V_{\mathrm{L}}
      \left(
        1-\gamma^{5}
      \right)
    \right]/2
  \right\}
  \psi = 0,
\ee
where $\gamma^{\mu}= \left( \gamma^{0},\bm{\gamma} \right)$, $\gamma^{5}=\mathrm{i}\gamma^{0}\gamma^{1}\gamma^{2}\gamma^{3}$,
and $\bm{\Sigma}=\gamma^{0}\bm{\gamma}\gamma^{5}$ are the Dirac matrices,
$P^{\mu}=\mathrm{i}\partial^{\mu}+eA^{\mu}$, $A^{\mu}=\left(0,0,Bx,0\right)$
is the vector potential, and $e>0$ is the absolute value of the elementary
charge. The effective potentials of the electroweak interaction $V_{\mathrm{R,L}}$
have the form~\cite{DvoSem15a},
\be\label{eq:VRL}
  V_{\mathrm{R}}= -\frac{G_{\mathrm{F}}}{\sqrt{2}}
  \left[
    n_{n}-n_{p}(1-4\xi)
  \right]
  2\xi,
  \quad
  V_{\mathrm{L}}= -\frac{G_{\mathrm{F}}}{\sqrt{2}}
  \left[
    n_{n}-n_{p}(1-4\xi)
  \right](2\xi-1),
\ee
where $n_{n,p}$ are the number densities of neutrons and protons,
$G_{\mathrm{F}}=1.17\times10^{-5}\,\text{GeV}^{-2}$ is the Fermi
constant, and $\xi=\sin^{2}\theta_{\mathrm{W}}\approx0.23$ is the
Weinberg parameter.

The solution of Eq.~\ref{eq:Direq} has the form~\cite{BalStuTok12},
\be\label{eq:psiCi}
 \psi^{\mathrm{T}} = \exp
  \left(
    -\mathrm{i}Et+\mathrm{i}p_{y}y+\mathrm{i}p_{z}z
  \right)
  \left(
    C_{1}u_{\mathrm{n}-1}, \mathrm{i}C_{2}u_{\mathrm{n}},
    C_{3}u_{\mathrm{n}-1},\mathrm{i}C_{4}u_{\mathrm{n}}
  \right)
\ee
where
\be
  u_{\mathrm{n}}(\eta) =
  \left(
    \frac{eB}{\pi}
  \right)^{1/4} 
  \exp
  \left(
    -\frac{\eta^{2}}{2}
  \right)
  \frac{H_{\mathrm{n}}(\eta)}{\sqrt{2^{\mathrm{n}}\mathrm{n}!}},
  \quad
  \mathrm{n} = 0,1,\dotsc,
\ee
are the Hermite functions, $H_{\mathrm{n}}(\eta)$
are the Hermite polynomials, $\eta=\sqrt{eB}x + p_{y} / \sqrt{eB}$,
$C_{i}$ are the spin coefficients, $i=1,\dots,4$, and $-\infty<p_{y,z}<+\infty$.

Using Eqs.~\ref{eq:Direq} and~\ref{eq:psiCi}, we get that the spin coefficients $C_{i}$ obey the system of equations,
\begin{align}\label{eq:Cisys}
  \left(
    E-\bar{V}-p_{z}+V_{5}
  \right)
  C_{1} - \sqrt{2eB\mathrm{n}}C_{2} +
  \left(
    m+\mu B
  \right)
  C_{3} = & 0,
  \notag
  \\
  \sqrt{2eB\mathrm{n}} C_{1} -
  \left(
    E-\bar{V}+p_{z}+V_{5}
  \right)
  C_{2} -
  \left(
    m-\mu B
  \right)
  C_{4} = & 0,
  \notag
  \\
  \left(
    m + \mu B
  \right)C_{1} +
  \left(
    E-\bar{V}+p_{z}-V_{5}
  \right)
  C_{3} + \sqrt{2eB\mathrm{n}}C_{4}	= & 0,
  \notag
  \\
  \left(
    m-\mu B
  \right)
  C_{2} + \sqrt{2eB\mathrm{n}}C_{3} +
  \left(
    E-\bar{V}-p_{z}-V_{5}
  \right)C_{4} = & 0,
\end{align}
where $\bar{V}=\left(V_{\mathrm{L}}+V_{\mathrm{R}}\right)/2$,
and $V_{5}=\left(V_{\mathrm{L}}-V_{\mathrm{R}}\right)/2$. To derive Eq.~\ref{eq:Cisys} we choose the Dirac matrices in the chiral representation~\cite{ItzZub80},
\be\label{eq:chirrep}
  \gamma^\mu =
  \begin{pmatrix}
    0 & -\sigma^\mu \\
    -\bar{\sigma}^\mu & 0 \ 
  \end{pmatrix},
  \quad
  \sigma^\mu = (\sigma_0, - \bm{\sigma}),
  \quad
  \bar{\sigma}^\mu = (\sigma_0, \bm{\sigma}),
\ee
where $\sigma_0$ is the unit $2\times 2$ matrix and $\bm{\sigma}$ are the Pauli matrices.

Equating the determinant of the system in Eq.~\ref{eq:Cisys} to zero, we get the energy levels $E$ for $\mathrm{n}>0$ in the form~\cite{BalStuTok12},
\begin{align}\label{eq:En}
  E = & \bar{V}+\mathcal{E},  
  \quad
  \mathcal{E}=\sqrt{p_{z}^{2}+m^{2}+2eB\mathrm{n} +
  \left(
    \mu B
  \right)^{2} + V_{5}^{2}+2sR^{2}},
  \notag
  \\
  R^{2} = & \sqrt{
  \left(
    p_{z} V_{5} - \mu B m
  \right)^{2} +
  2eB\mathrm{n}
  \left[
    V_{5}^{2} +
    \left(
      \mu B
    \right)^{2}
  \right]},
\end{align}
where $s=\pm1$ is the discrete spin quantum number.
 
To determine $C_i$ at $\mathrm{n}>0$ we notice that the spin operator~\cite{BalStuTok12}
\be
  \hat{S} =
  \frac{V_{5}\hat{S}_{\mathrm{long}} -
  \mu B\hat{S}_{\mathrm{tr}}}{\sqrt{V_{5}^{2}+(\mu B)^{2}}},  
  \quad
  \hat{S}_{\mathrm{long}} = \frac{(\bm{\Sigma}\mathbf{P})}{m},
  \quad
  \hat{S}_{\mathrm{tr}} =
  \Sigma_{3}-\frac{\mathrm{i}}{m}(\bm{\gamma}\times\mathbf{P})_{3}
\ee
commutes with the Hamiltonian of Eq.~\ref{eq:Direq}. Basing on the fact that the wave function in Eq.~\ref{eq:psiCi} is the eigenfunction of  the operator $\hat{S}$, it is convenient to represent $C_i$ in terms of the new auxiliary coefficients $\mathcal{A}$
and $\mathcal{B}$ as
\begin{align}\label{eq:CAB}
  \begin{pmatrix}
    C_{1} \\
    C_{3}
  \end{pmatrix} = &
  \frac{1}{\sqrt{2}}
  \sqrt{1-\frac{s}{R^2}(p_{z}V_{5}-\mu B m)}
  \begin{pmatrix}
    Z & -\mu B/Z \\
    \mu B/Z & Z
  \end{pmatrix}
  \begin{pmatrix}
    \mathcal{A}\\
    \mathcal{B}
  \end{pmatrix},
  \notag
  \\
  \begin{pmatrix}
    C_{2} \\
    C_{4} 
  \end{pmatrix} = &
  \frac{s}{\sqrt{2}}
  \sqrt{1+\frac{s}{R^2}(p_{z}V_{5}-\mu B m)}
  \begin{pmatrix}
    Z & \mu B/Z \\
    -\mu B/Z & Z \ 
  \end{pmatrix}
  \begin{pmatrix}
    \mathcal{A} \\
    \mathcal{B}
  \end{pmatrix},
\end{align}
where $Z = \sqrt{V_{5} + \sqrt{V_{5}^2 + (\mu B)^2}}$.

Inserting Eq.~\ref{eq:CAB} to Eq.~\ref{eq:Cisys}, we get that $\mathcal{A}$
and $\mathcal{B}$ are completely defined by the following relation:
\begin{gather}
  \mathcal{A}^{2} =
  \left\{
    1-\frac{s R^2+(\mu B)^{2}+V_{5}^{2}}{\mathcal{E}\sqrt{V_{5}^{2}+(\mu B)^{2}}}
  \right\}
  \mathcal{C}^{2},
  \quad
  \mathcal{B}^{2}	=
  \left\{
    1+\frac{s R^2+(\mu B)^{2}+V_{5}^{2}}{\mathcal{E}\sqrt{V_{5}^{2}+(\mu B)^{2}}}
  \right\}
  \mathcal{C}^{2},
  \notag
  \\
  \mathcal{A}\mathcal{B} =
  -\frac{mV_{5}+\mu Bp_{z}}{\mathcal{E}\sqrt{V_{5}^{2}+(\mu B)^{2}}}
  \mathcal{C}^{2}.
  \label{eq:ABAB}
\end{gather}
The coefficient $\mathcal{C}$ can be found if we normalize the wave function $\psi$ as
\be
  \int\mathrm{d}^{3}x
  \psi_{p_{y}p_{z}\mathrm{n}}^{\dagger}\psi_{p'_{y}p'_{z}\mathrm{n}'} =
  \delta\left(p_{y}-p'_{y}\right)\delta\left(p_{z}-p'_{z}\right)\delta_{\mathrm{nn}'}.
\ee
In this situation, the spin coefficients obey the relation,
\be\label{eq:Cinorm}
  \sum_{i=1}^{4}|C_{i}|^{2}=\frac{1}{(2\pi)^{2}},
\ee
at any $\mathrm{n} \geq 0$. Finally, we get that
\be\label{eq:C}
  \mathcal{C}^{2}=\frac{1}{4(2\pi)^{2}\sqrt{V_{5}^{2}+(\mu B)^{2}}}.
\ee 
We can see that Eqs.~\ref{eq:CAB}, \ref{eq:ABAB}, and~\ref{eq:C} completely define the spin coefficients $C_i$ at $\mathrm{n}>0$.

At $\mathrm{n}=0$, both the energy spectrum and the spin coefficients can be found directly from Eq.~\ref{eq:Cisys} since, in this case, the electron spin
has only one direction and hence $C_{1}=C_{3}=0$. Thus, the energy spectrum reads
\be\label{eq:E0}
  E = \bar{V} + \mathcal{E},
  \quad
  \mathcal{E} = \sqrt{
  \left(
    p_{z}+V_{5}
  \right)^{2} +
  \left(
    m-\mu B
  \right)^{2}}.
\ee
Using Eq.~\ref{eq:Cinorm}, we obtain that the nonzero spin coefficients $C_{2,4}$ have the form,
\be\label{eq:C2C4}
  |C_2|^2 = \frac{1}{2(2\pi)^2\mathcal{E}}\frac{(m-\mu B)^2}{(\mathcal{E}+p_z+V_5)},
  \quad
  |C_4|^2 = \frac{\mathcal{E}+p_z+V_5}{2(2\pi)^2\mathcal{E}}.
\ee
Note that, while solving Eq.~\ref{eq:Direq}, we take into account only electron rather than positrons degrees of freedom.

\section{Calculation of the electric current\label{sec:CURR}}

Using the exact solution of the Dirac equation, which is found above, we can calculate the
electric current of electrons in this matter. This current has the
form~\cite{Vil80},
\be\label{eq:Jzgen}
  \mathbf{J} =
  - e
  \sum_{\mathrm{n}=0}^{\infty}
  \sum_{s}
  \int_{-\infty}^{+\infty}
  \mathrm{d}p_{y}\mathrm{d}p_{z}
  \bar{\psi}\bm{\gamma}\psi f(E-\chi),
\ee
where $f(E)=\left[\exp(\beta E)+1\right]^{-1}$ is the Fermi-Dirac
distribution function, $\beta=1/T$ is the reciprocal temperature,
and $\chi$ is the chemical potential. First, we notice that $J_{x,y}\sim\bar{\psi}\gamma^{1,2}\psi=0$
because of the orthogonality of Hermite functions with different indexes. Hence, only $J_{z}\sim\bar{\psi}\gamma^{3}\psi$ should be considered. Then, using Eq.~\ref{eq:CAB}, we can derive the identity
\begin{align}\label{eq:dpy}
  \int_{-\infty}^{+\infty}
  \mathrm{d}p_{y}\psi^{\dagger}\gamma^0\gamma^3\psi = &
  eB
  \left(
    |C_{1}|^{2}+|C_{4}|^{2}-|C_{2}|^{2}-|C_{3}|^{2}
  \right)
  \notag
  \\
  & =
  -2 eB
  \left[
    4\mu B\mathcal{AB} +
    s\frac{V_{5}}{R^2}(p_{z}V_{5}-\mu B m)
    \left(
      \mathcal{A}^{2}-\mathcal{B}^{2}
    \right)
  \right],
\end{align}
which is valid in the chiral representation of Dirac matrices in Eq.~\ref{eq:chirrep}.

Now let us consider the contribution of the energy levels, with $\mathrm{n}>0$,
to $J_{z}$. Basing on Eqs.~\ref{eq:ABAB}, \ref{eq:C}, and~\ref{eq:dpy}, we obtain that it has the form,
\be\label{eq:Jzn>0}
  J_{z}^{(\mathrm{n}>0)} =
  -\frac{e^{2}B}{(2\pi)^{2}}
  \sum_{\mathrm{n}=1}^{\infty}
  \sum_{s=\pm1}
  \int_{-\infty}^{+\infty}
  \frac{\mathrm{d}p_{z}}{\mathcal{E}}
  \left[
    p_{z}
    \left(
      1+s\frac{V_{5}^{2}}{R^{2}}
    \right) -
    s\frac{\mu BmV_{5}}{R^{2}}
  \right]f(E-\chi).
\ee
To find the first nonzero term in Eq.~\ref{eq:Jzn>0} we decompose $J_{z}^{(\mathrm{n}>0)}$ in a series in
$\mu B$ and $V_{5}$. Finally, we get that
\be\label{eq:JznV5mu}
  J_{z}= \mu mV_{5}B^{2}\frac{e^{2}}{\pi^{2}}
  \sum_{\mathrm{n}=1}^{\infty}
  \int_{0}^{+\infty}
  \frac{\mathrm{d}p}{\mathcal{E}_{\mathrm{eff}}^{2}}
  \left[
    \left(
      1-\frac{3p^{2}}{\mathcal{E}_{\mathrm{eff}}^{2}}
    \right)
    \left(
      f'-\frac{f}{\mathcal{E}_{\mathrm{eff}}}
    \right) +
    \frac{p^{2}}{\mathcal{E}_{\mathrm{eff}}}f''
  \right],
\ee
where $\mathcal{E}_{\mathrm{eff}}=\sqrt{p^{2}+m_{\mathrm{eff}}^{2}}$
and $m_{\mathrm{eff}}=\sqrt{m^{2}+2eB\mathrm{n}}$. The argument of the distribution
function in Eq.~\ref{eq:JznV5mu} is $\mathcal{E}_{\mathrm{eff}}+\bar{V}-\chi$.

As an example, we shall consider a strongly degenerate electron gas. In
this situation, $f=\theta(\chi-\bar{V}-\mathcal{E}_{\mathrm{eff}})$,
where $\theta(z)$ is the Heaviside step function. We can also disregard
the positrons contribution to $J_{z}$. The direct calculation of
the current in Eq.~\ref{eq:JznV5mu} gives
\be\label{eq:Jzdeggen}
  J_{z}= - 2\mu mV_{5}B^{2}
  \frac{e^{2}}{\pi^{2}
  \tilde{\chi}^{3}}
  \sum_{\mathrm{n}=1}^{\infty}
  \sqrt{\tilde{\chi}^{2}-m_{\mathrm{eff}}^{2}}\theta
  \left(
    \tilde{\chi}-m_{\mathrm{eff}}
  \right),
\ee
where $\tilde{\chi}=\chi-\bar{V}$. One can see that $J_{z}$ in Eq.~\ref{eq:Jzdeggen}
is nonzero if $B<\tilde{B}$, where $\tilde{B}=\left(\tilde{\chi}^{2}-m^{2}\right)/2e$.
If the magnetic field is relatively strong and is close to $\tilde{B}$,
then only the first energy level with $\mathrm{n}=1$ contributes
to $J_{z}$, giving one
\be
  J_{z} = - \frac{8 \alpha_{\mathrm{em}}}{\pi}
  \frac{\mu B mV_{5}}{\tilde{\chi}^{3}} B
  \sqrt{\tilde{\chi}^{2}-m^{2}-2eB} \to 0,
\ee
where $\alpha_{\mathrm{em}}=e^{2}/4\pi$. In the opposite situation,
when $B\ll\tilde{B}$, one gets that
\be
  J_{z} = - \frac{8\alpha_{\mathrm{em}}}{3\pi e}
  \frac{\mu B mV_{5}}{\tilde{\chi}^{3}}
  \left(
    \tilde{\chi}^{2}-m^{2}-2eB
  \right)^{3/2} 
  \approx
  - \frac{8\alpha_{\mathrm{em}}}{3\pi e} \mu m V_{5} B,
\ee
i.e. the current is proportional to the magnetic field strength.

Finally, let us consider the contribution of the lowest energy level $\mathrm{n}=0$ to $J_z$. Using Eqs.~\ref{eq:E0}, \ref{eq:C2C4}, and~\ref{eq:dpy}, we rewrite $J_{z}^{(\mathrm{n}=0)}$ in Eq.~\ref{eq:Jzgen} as
\begin{align}\label{eq:Jz0}
  J_{z}^{(\mathrm{n}=0)} =	& e^{2}B
  \int_{-\infty}^{+\infty}\mathrm{d}p_{z}
  \left(
    |C_{2}|^{2}-|C_{4}|^{2}
  \right)f(E-\chi)
  \notag
  \\
  & =
  - \frac{e^{2}B}{(2\pi)^{2}}
  \int_{-\infty}^{+\infty}\mathrm{d}p_{z}
  \frac{p_{z}+V_{5}}{\sqrt{
  \left(
    p_{z}+V_{5}
  \right)^{2}+
  \left(
    m-\mu B
  \right)^{2}}}f
  \left(
    E-\chi
  \right)=0,
\end{align}
where we take into account the expression for $E$ in Eq.~\ref{eq:E0}. Eq.~\ref{eq:Jz0} means that the lowest energy level with $\mathrm{n}=0$ does not contribute to the electric current along the magnetic field. This result extends our recent finding~\cite{Dvo16a} to the situation when the anomalous magnetic moment is accounted for. Analogously to Eq.~\ref{eq:Jz0}, one can show that positrons do not contribute to the current either. Note that the result is Eq.~\ref{eq:Jz0} is valid for arbitrary characteristics of plasma, external fields, as well as the mass and the magnetic moment of an electron.

It is interesting to compare the appearance of the new current along the magnetic field in Eq.~\ref{eq:JznV5mu} with CME~\cite{MirSho15,Kha15}. Vilenkin~\cite{Vil80} showed that only massless electrons at the zero Landau level in an external magnetic field contribute to the generation of the anomalous current along the magnetic field. This feature remains valid in the presence of the background electroweak matter~\cite{DvoSem15a,DvoSem15b}. The current of such massless particles is exited since, at the zero Landau level, left electrons move along the magnetic field, whereas right particles move in the opposite direction~\cite{DvoSem15a,DvoSem15b}. Electrons at higher Landau levels can move arbitrarily with respect to the magnetic field. Therefore, if one has a different population of left and right electrons at the lowest Landau level, there is a nonzero current is the system $\mathbf{J}_{5} \sim \left( \mu_{\mathrm{R}}-\mu_{\mathrm{L}} \right) \mathbf{B}$, which is the manifestation of CME.

In the situation described in the present work, i.e. when massive electrons with nonzero anomalous magnetic moment move in the electroweak matter, the particles at the lowest energy level can move in any direction with respect to the magnetic field, i.e. $-\infty < p_z < +\infty$. Moreover, there is no asymmetry for electrons with $p_z > 0$ and $p_z <0$. It results from Eq.~\ref{eq:E0} if we replace $p_z \to p_z - V_5$ there. On the contrary, higher energy levels with $\mathrm{n} > 0$ in Eq.~\ref{eq:En} are not symmetric with respect to the transformation $p_z \to - p_z$. The reflectional symmetry cannot be restored by any replacement of $p_z$. Therefore electrons having $p_z > 0$ and $p_z < 0$ will have different energies and hence different velocities $v_z = p_z / \mathcal{E}$. The electric current along the magnetic field $\mathbf{B} = B \mathbf{e}_z$ is proportional to $\langle v_z \rangle$. Thus such a current should be nonzero, with only higher energy levels contributing to it. It is interesting to mention that the term in Eq.~\ref{eq:En} which violates the reflectional symmetry $p_z \to - p_z$ is proportional to $\mu B m V_5$. It is this factor which $J_z$ in Eq.~\ref{eq:JznV5mu} is proportional to.

\section{Instability of the magnetic field in NS\label{sec:INST}}

Returning to the vector notations we get the current in Eq.~\ref{eq:Jzdeggen} takes the form,
\be\label{eq:JPiB}
  \mathbf{J} = \Pi\mathbf{B},
  \quad
  \Pi=-8\mu mV_{5}B\frac{\alpha_{\mathrm{em}}}{\pi\tilde{\chi}^{3}}
  \sum_{\mathrm{n}=1}^{N}
  \sqrt{\tilde{\chi}^{2}-m_{\mathrm{eff}}^{2}},
\ee
where $N$ is maximal integer, for which $\tilde{\chi}^{2}-m^{2}-2eBN\geq0$.
To study the magnetic field evolution in the presence of the additional
current in Eq.~\ref{eq:JPiB} we take this current into account
in the Maxwell equations along with the usual ohmic current $\mathbf{J}=\sigma_{\mathrm{cond}}\mathbf{E}$,
where $\sigma_{\mathrm{cond}}$ is the matter conductivity and $\mathbf{E}$
is the electric field. Considering the magnetohydrodynamic approximation,
which reads $\sigma_{\mathrm{cond}}\gg\omega$, where $\omega$ is
the typical frequency of the electromagnetic fields variation, we
derive the modified Faraday equation for the magnetic field evolution,
\begin{align}\label{eq:mFe}
  \frac{\partial\mathbf{B}}{\partial t} = &
  \frac{\Pi}{\sigma_{\mathrm{cond}}}
  \left(
    \nabla \times \mathbf{B}
  \right) +
  \frac{1}{\sigma_{\mathrm{cond}}}\nabla^{2}\mathbf{B}
  \notag
  \\
  & +
  \frac{1}{\sigma_{\mathrm{cond}}B}\frac{\mathrm{d}\Pi}{\mathrm{d}B}
  \left[
    B^{2}
    \left(
      \nabla\times\mathbf{B}
    \right)-
    \mathbf{B}
    \left(
      \mathbf{B}\cdot\nabla\times\mathbf{B}
    \right)-
    \left(
      \mathbf{B}\times
      \left(
        \mathbf{B}\nabla
      \right)\mathbf{B}
    \right)
  \right],
\end{align}
where we neglect the coordinate dependence of $\sigma_{\mathrm{cond}}$.

Let us consider the evolution of the magnetic field given by the Chern-Simons
wave, with the amplitude $A(t)$, corresponding to the maximal negative
helicity, $\mathbf{A}(z,t)=A(t)\big(\mathbf{e}_{x}\cos kz +\mathbf{e}_{y}\sin kz\big)$
or $\mathbf{B}(z,t)=B(t)\big(\mathbf{e}_{x}\cos kz+\mathbf{e}_{y}\sin kz\big)$,
where $k=1/L$ is the wave number determining the length scale of
the magnetic field $L$ and $B(t)=-kA(t)$. In this situation, Eq.~\ref{eq:mFe} can be
simplified. The equation for the amplitude of the magnetic field $B$ takes the form,
\be\label{eq:dotB}
  \dot{B}=-\frac{k}{\sigma_{\mathrm{cond}}}
  \left(
    k+\Pi
  \right)B.
\ee
Since $\Pi$ in Eq.~\ref{eq:JPiB} is negative, the magnetic field,
described by Eq.~\ref{eq:dotB}, can be unstable.

We shall apply Eq.~\ref{eq:JPiB} to describe the magnetic field
amplification in a dense degenerate matter which can be found in NS.
In this situation, $n_{n}=1.8\times10^{38}\,\text{cm}^{-3}$ and $n_{p}\ll n_{n}$.
Using Eq.~\ref{eq:VRL} for this number density of neutrons, one
gets that $V_{5} = G_\mathrm{F} n_n /2\sqrt{2} = 6\,\text{eV}$. The number density of electrons can
reach several percent of the nucleon density in NS. We shall take
that $n_{e}=9\times10^{36}\,\text{cm}^{-3}$, which gives one $\chi = (3\pi^2 n_e)^{1/3} = 125\,\text{MeV}$~\cite{DvoSem15b}.
Thus electrons are ultrarelativistic and we can take that $\tilde{\chi}\approx\chi$.
We shall study the magnetic field evolution in NS in the time interval
$t_{0}<t<t_{\mathrm{max}}$, where $t_{0}\sim10^{2}\,\text{yr}$ and
$t_{\mathrm{max}}\sim10^{6}\,\text{yr}$. In this time interval, NS
cools down from $T_{0}\sim10^{8}\,\text{K}$ mainly by the neutrino
emission~\cite{YakPet04}. In this situation, the matter conductivity
in Eq.~\ref{eq:dotB} becomes time dependent $\sigma_{\mathrm{cond}}(t)=\sigma_{0}(t/t_{0})^{1/3}$~\cite{DvoSem15b},
where $\sigma_{0}=2.7\times10^{5}\,\text{GeV}$. Here we use the chosen
electron density.

We shall discuss the amplification of the seed magnetic field $B_{0}=10^{12}\,\text{G}$, which is typical for a young pulsar.
In such strong magnetic fields, the anomalous magnetic moment of an
electron was found by Ternov et al.~\cite{Ter69} to depend on the magnetic
field strength. We can approximate $\mu$ as
\be\label{eq:mu}
  \mu = \frac{e}{2m}\frac{\alpha_{\mathrm{em}}}{2\pi}
  \left(
    1-\frac{B}{B_{c}}
  \right),
\ee
where $B_{c}=m^{2}/e=4.4\times10^{13}\,\text{G}$. Note that Eq.~\ref{eq:mu}
accounts for the change of the sign of $\mu$ at $B\approx B_{c}$
predicted by Ternov et al.~\cite{Ter69}.

\begin{figure}
  \centering
  \subfigure[]
  {\label{1a}
  \includegraphics[scale=.11]{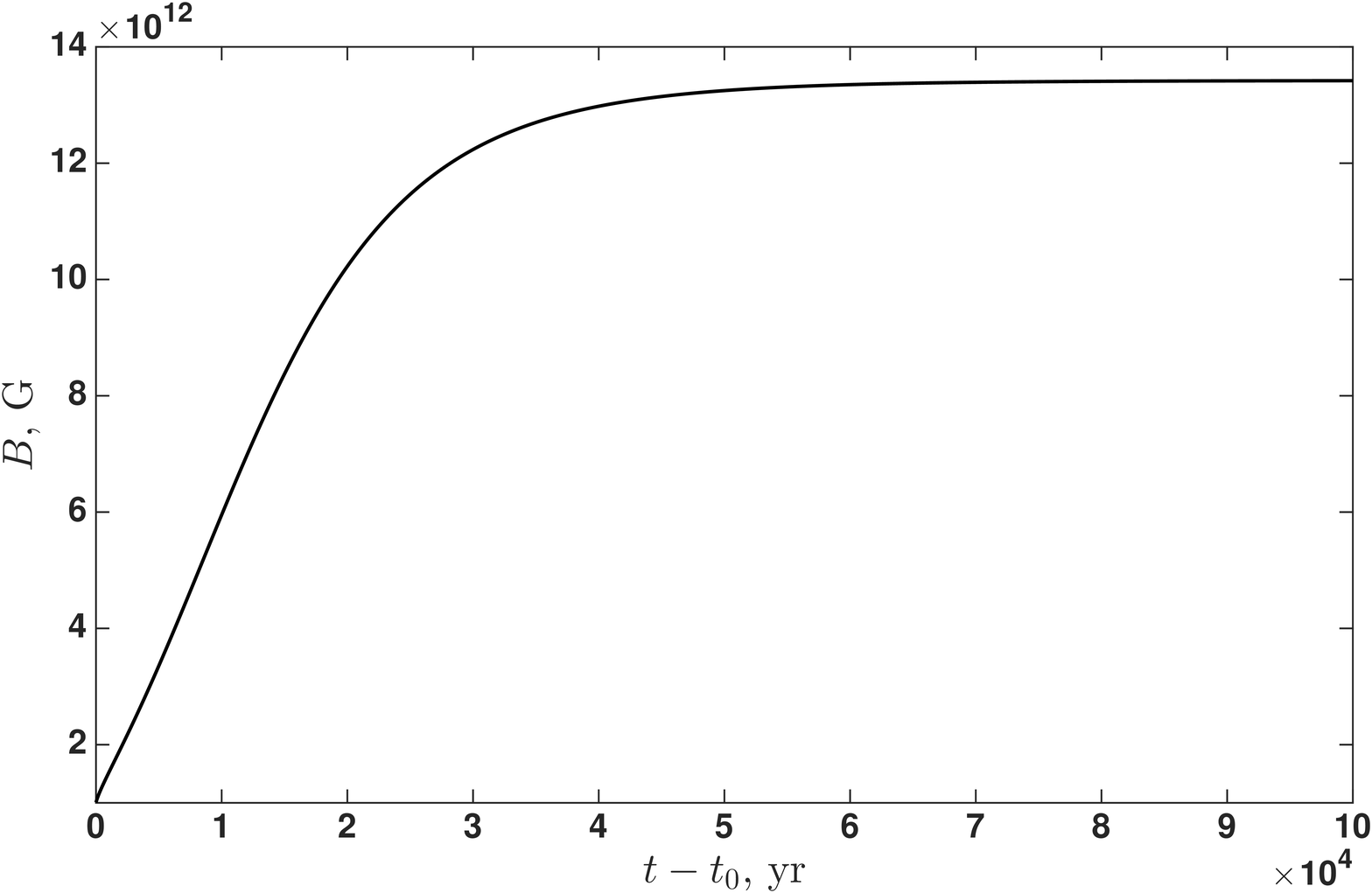}}
  \hskip-.9cm
  \subfigure[]
  {\label{1b}
  \includegraphics[scale=.11]{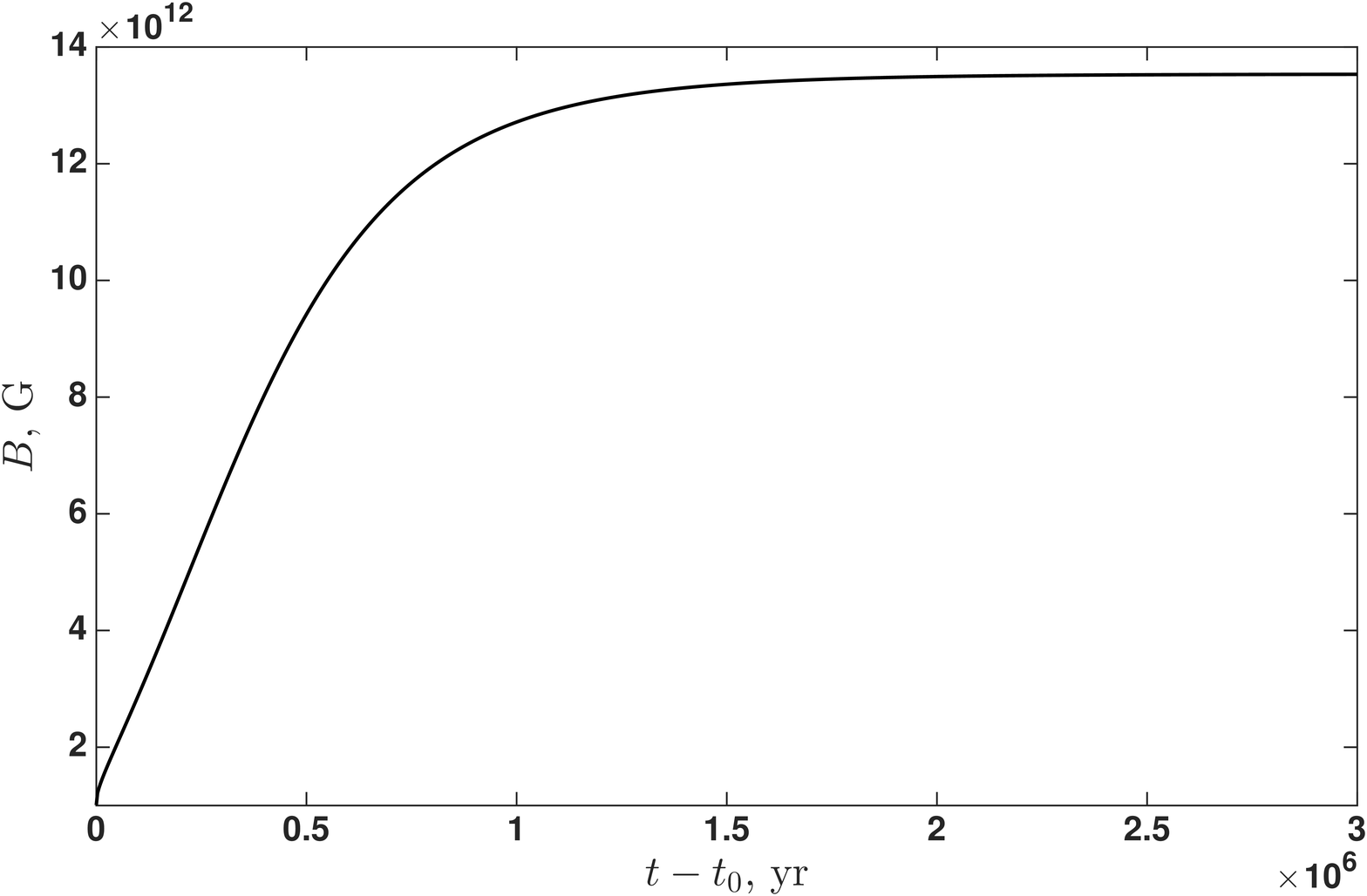}}
  \protect
  \caption{Magnetic field evolution obtained by the numerical solution of  
  Eq.~\ref{eq:dotB} for different length scales.
  (a) $L=10^{2}\,\text{cm}$, and (b) $L=10^{3}\,\text{cm}$.
  \label{fig:Bfield}}
\end{figure}

The evolution of the magnetic field for the chosen initial conditions
is shown in Fig.~\ref{fig:Bfield} for different length scales. One
can see that, if the magnetic
field is enhanced from $B_{0}=10^{12}\,\text{G}$, it reaches the saturated strength $B_{\mathrm{sat}} \approx 1.3\times10^{13}\,\text{G}$.
Thus, both quenching factors in Eqs.~\ref{eq:JPiB} and~\ref{eq:mu}
are important. One can see in Fig.~\ref{fig:Bfield} that a larger
scale magnetic field grows slower. The further enhancement of the
magnetic field scale compared to $L=10^{3}\,\text{cm}$ corresponding
to Fig.~\ref{1b} is inexpedient since the growths time
would significantly exceed $10^{6}\,\text{yr}$. At such long evolution
times, NS cools down by the photon emission from the stellar surface
rather than by the neutrino emission~\cite{YakPet04}.

The energy source, powering the magnetic field growth shown in Fig.~\ref{fig:Bfield},
can be the kinetic energy of the stellar rotation. To describe the
energy transmission from the rotational motion of matter to the magnetic
field one should take into account the advection term $\nabla\left(\mathbf{v}\times\mathbf{B}\right)$
in the right hand side of Eq.~\ref{eq:mFe}. Here $\mathbf{v}$
is the matter velocity. Moreover one should assume the differential
rotation of NS~\cite{Sha00}. For this purpose we should take that
NS is not in a superfluid state. This case is not excluded by the
observational data~\cite{GneYakPot01}. We have estimated the spin
down of NS with the radius $R\sim10\,\text{km}$ and the initial rotation
period $P_{0}\sim10^{-3}\,\text{s}$ basing on the fact that
the total energy, $I\Omega^{2}/2+B^{2}V/2$, is constant. Here $I$
is the moment of inertia of NS, $\Omega$ is the angular velocity,
and $V$ is the NS volume. For $B_{\mathrm{sat}}\approx1.3\times10^{13}\,\text{G}$
shown in Fig.~\ref{fig:Bfield}, the relative change of the period
is $(P-P_{0})/P_{0}\sim10^{-9}$. Hence only a small fraction of the initial rotational energy is transmitted to the energy of a growing magnetic field.

The obtained results can be used for the explanation of electromagnetic
flashes emitted by magnetars~\cite{TurZanWat15}. Beloborodov \& Levin~\cite{BelLev14} suggested that magnetar bursts,
happening in the stellar magnetosphere, are triggered by plastic deformations of the magnetar crust driven
by a thermoplastic wave (TPW). TPW can be excited by a fluctuation of the internal magnetic field with the length scale of about several meters~\cite{LiLevBel16} having the strength $B \gtrsim 10^{13}\,\text{G}$~\cite{Lan16}. As one can see in Fig.~\ref{fig:Bfield},
these conditions are fulfilled in our case. Therefore the instability of the magnetic field predicted in our model can excite TPW which then causes a magnetar burst.

\section{Conclusion}

In conclusion we mention that, in the present work, we have considered
the generation of the electric current of charged fermions, e.g.,
electrons, flowing along the external magnetic field. This current
is nonzero if electrons electroweakly interact with background matter
as well as if the nonzero mass and the nonzero anomalous magnetic
moment are accounted for. Unlike the situation of massless fermions,
when, owing to CME, $\mathbf{J}_{5}\sim\mathbf{B}$ is created by
the polarization effects at the lowest energy level~\cite{DvoSem15a,DvoSem15b},
in our case, only higher energy levels with $\mathrm{n}>0$ contribute
to the current. We also mention that the role of a nonzero anomalous
magnetic moment is crucial since, as found by Dvornikov~\cite{Dvo16a},
the current of massive charged particles electroweakly interacting
with background matter is vanishing at any $\mathrm{n}\geq0$.

We have revealed that a magnetic field turns out to be unstable if
the new current in Eq.~\ref{eq:JznV5mu} is taken into account.
As an example of the obtained results, we have discussed the enhancement
of the magnetic field in a dense degenerate matter. Using the background
matter with characteristics typical to NS, we have obtained the amplification
of the seed field $B_{0}=10^{12}\,\text{G}$ by more than one order
of magnitude. The generated magnetic field has a relatively small scale
$L\sim(10^{2}-10^{3})\,\text{cm}$. The time for the field growth
is $(10^{5}-10^{6})\,\text{yr}$ depending on the length scale. Finally
we have considered the implication of our results for the explanation
of magnetar bursts.

\section*{Acknowledgments}

I am thankful to the organizers of the 52$^\text{nd}$ Rencontres de Moriond for the invitation and a financial support,
as well as to the Tomsk State University Competitiveness Improvement
Program and RFBR (research project No.~15-02-00293).



\end{document}